\documentclass[a4paper,12pt]{article}

\usepackage{amsmath}
\usepackage{amssymb}
\usepackage{amsfonts}
\usepackage{bbm} 
\usepackage{fleqn} 
\usepackage[footnotesize]{caption}
\usepackage{mathrsfs}
\usepackage{graphicx} 
\usepackage[center,footnotesize,hang]{subfigure}

\def\<{\left\langle}
\def\>{\right\rangle}
\def\ChargeC{\mathrm{C}}

\def\SU{\text{SU}}
\def\SO{\text{SO}}

\def\simlt{\stackrel{<}{{}_\sim}}

\def\be{\begin{equation}}
\def\ee{\end{equation}}
\def\beq{\begin{equation}}
\def\eeq{\end{equation}}
\def\bea{\begin{eqnarray}}
\def\eea{\end{eqnarray}}

\def\da{{\delta}_1}
\def\db{{\delta}_2}
\def\dc{{\delta}_3}
\def\daP{{\delta}'_1}
\def\dbP{{\delta}'_2}
\def\dcP{{\delta}'_3}

\newcommand{\CenterEps}[2][1]{\ensuremath{\vcenter{\hbox{\includegraphics[scale=#1]{#2.eps}}}}}

\def\<{\left\langle}
\def\>{\right\rangle}
\def\ChargeC{\mathrm{C}}

\begin{document}

\newcommand{\sheptitle}
{{\Large Sequential Dominance}}

\newcommand{\shepauthor}
{Stefan Antusch\footnote{E-mail: \texttt{santusch@hep.phys.soton.ac.uk}} and 
S. F. King\footnote{E-mail: \texttt{sfk@hep.phys.soton.ac.uk}} }

\newcommand{\shepaddress}
{School of Physics and Astronomy,
University of Southampton, \\ Southampton, SO17 1BJ, U.K.}

\newcommand{\shepabstract}
{We review the mechanism of sequential right-handed neutrino 
dominance proposed in the
framework of the type I see-saw mechanism to account for bi-large
neutrino mixing and a neutrino mass hierarchy in a natural way. 
We discuss how sequential dominance may also be applied to the
right-handed charged leptons, which alternatively 
allows bi-large lepton mixing from the charged lepton sector. We review how
such sequential dominance models may be upgraded to include type II
see-saw contributions, resulting in a partially degenerate neutrino
mass spectrum with bi-large lepton mixing arising from sequential
dominance. We also summarise the model building applications and the
phenomenological implications of sequential dominance.}

\begin{titlepage}
\begin{flushright}
hep-ph/0405272\\
\end{flushright}
\begin{center}
{\large{\bf \sheptitle}}
\bigskip \\ \shepauthor \\ \mbox{} \\ {\it \shepaddress} \\ \vspace{.5in}
{\bf Abstract} \bigskip \end{center} \setcounter{page}{0}
\shepabstract
\end{titlepage}

\section{Introduction}

Neutrino masses and mixing angles must now be regarded
as unavoidable consequences of the 
firmly established atmospheric and solar neutrino oscillation experiments
\cite{King:2003jb}. A profound consequence of this is that the
minimal Standard Model must necessarily be incomplete, and must be
extended in some way to account for neutrino masses and mixings.
The simplest way to do this appears to be to add right-handed
neutrinos to the Standard Model. Since the right-handed neutrinos
are gauge singlets electroweak symmetry does not
prevent then from having large Majorana masses ranging from a few TeV
up to the Planck scale. The right-handed neutrinos
may also couple to left-handed leptons via the usual Higgs doublets.
The combination of very large right-handed neutrino Majorana masses,
and weak scale Dirac masses from the Higgs couplings leads to
suppressed left-handed Majorana neutrino masses which may be identified
with the physical neutrino masses responsible for atmospheric and solar
neutrino oscillations. This scenario, proposed some time
ago in \cite{seesaw}, is known as the see-saw mechanism.

Given the simplicity of the see-saw mechanism it has been widely
applied to understanding the pattern of neutrino masses and mixings
implied by the atmospheric and solar neutrino oscillation data 
\cite{King:2003jb}. Although there are alternatives to the see-saw
mechanism involving large extra dimensions \cite{Arkani-Hamed:1998vp}
or R-parity violating supersymmetry \cite{Hirsch:2004he}, we shall
consider only the see-saw mechanism here, although we shall
later consider a more complicated version of the see-saw mechanism 
called the type II see-saw mechanism which also involves heavy Higgs
triplets (see e.g.~\cite{Lazarides:1980nt}). We shall also
only consider the case of three active neutrinos, which is 
the minimal case consistent with the confirmed 
atmospheric and solar neutrino oscillation data.

Within the framework as described above the goal of the see-saw
mechanism is to account for large atmospheric neutrino mixing
($\theta_{23}\sim 45^\circ$, close to maximal), and
large solar neutrino mixing ($\theta_{12}\sim 30^\circ$, but not maximal) 
together with the observed atmospheric and solar neutrino mass squared
differences \cite{King:2003jb}. The solar data is consistent with
the so called large mixing angle (LMA) MSW solution 
\cite{MSW}. The third remaining mixing angle associated with
the three active neutrinos is so far unmeasured but must be quite
small ($\theta_{13}\simlt 13^\circ$) \cite{Apollonio:1999ae}.
The neutrino oscillation data does not determine the absolute
scale of neutrino masses, nor does it uniquely fix the ordering
of neutrino masses, however in a normally ordered hierarchical
scheme the neutrino mass values would be 
roughly given by $m_3\sim 0.05 \ {\rm eV}$,
$m_2\sim 0.008 \ {\rm eV}$, with $m_1\ll m_2$.
However $m_1$ could in principle 
be substantially larger, up to 
the cosmological limit of about 0.23 eV \cite{Elgaroy:2002bi}.

It has frequently been observed that the simultaneous appearance
of hierarchical neutrino masses and two large mixing angles
is not natural in the see-saw mechanism. An important exception to 
this is the sequential dominance mechanism 
\cite{King:1998jw,King:1999cm,King:1999mb,King:2002nf,King:2002qh} 
(see also \cite{Davidson:1998bi}) which is the subject of
this focus. Sequential dominance is not in itself a model, but is
a sub-mechanism within the general framework of the see-saw mechanism,
that may be applied to constructing different classes of models.
The starting point of sequential dominance is to assume
that one of the right-handed neutrinos
contributes dominantly in the see-saw mechanism to the heaviest
neutrino mass, with the atmospheric mixing angle being determined
by a simple ratio of two Yukawa couplings \cite{King:1998jw,King:1999cm}, 
which is sometimes referred
to as single right-handed neutrino dominance. 
Sequential dominance corresponds to the
further assumption that, together with single right-handed neutrino dominance, 
a second right-handed neutrino contributes
dominantly to the second heaviest neutrino mass, with the large
solar mixing angle interpreted
as a ratio of Yukawa couplings \cite{King:1999mb,King:2002nf}.
The third right-handed neutrino
is effectively decoupled from the see-saw mechanism, and plays
no part in determining the neutrino mass spectrum, although it may
play a cosmological role. If the decoupled right-handed neutrino is
also the heaviest one then 
sequential dominance is effectively equivalent to 
having two right-handed neutrinos \cite{King:1999cm,King:1999mb}.

We also review how sequential dominance may be generalized to include the
right-handed charged leptons \cite{Antusch:2004re}, 
which allows bi-large charged lepton
mixing consistent with a neutrino mass hierarchy. We then show how
such sequential dominance models may be upgraded to include type II
see-saw contributions \cite{Antusch:2004xd}, 
resulting in a partially degenerate neutrino
mass spectrum with bi-large lepton mixing arising from sequential
dominance.

In section 2 we recall the type I and type II
see-saw mechanisms. Section 3 shows how the
type I see-saw mechanism can lead to a hierarchical 
pattern of neutrino masses with bi-large neutrino mixing
in a natural way using sequential
right-handed neutrino dominance.
Section 4 shows how the type I see-saw mechanism can lead to
bi-large charged lepton mixing, with naturally small neutrino mixing
and hierarchical neutrino masses, 
using sequential dominance in the right-handed lepton sector.
Section 5 shows how a partially degenerate neutrino 
mass spectrum could originate from the type II see-saw
mechanism, with the neutrino mass splittings and mixings
controlled by sequential dominance.
Section 6 briefly reviews some of the model building applications,
while section 7 discusses the phenomenological implications 
of sequential dominance. Section 8 concludes the review.
Our conventions are stated in the Appendix.


\newpage

\section{The See-Saw Mechanism}
The see-saw mechanism provides a convincing explanation for the smallness 
of neutrino masses. 
In this section, we review its simplest form, the
type I see-saw mechanism and its generalization to  
the type II see-saw mechanism.

\subsection{Type I See-Saw}
Before discussing the see-saw mechanism it is worth first reviewing
the different types of neutrino mass that are possible. So far we
have been assuming that neutrino masses are Majorana masses of the form
\beq
m^\nu_{LL}\overline{\nu_\mathrm{L}} \nu_\mathrm{L}^\ChargeC
\label{mLL}
\eeq
where $\nu_\mathrm{L}$ is a left-handed neutrino field and $\nu_\mathrm{L}^\ChargeC$ is
the CP conjugate of a left-handed neutrino field, in other words
a right-handed antineutrino field. Such Majorana masses are possible
to since both the neutrino and the antineutrino
are electrically neutral and so
Majorana masses are not forbidden by electric charge conservation.
For this reason a Majorana mass for the electron would
be strictly forbidden. However such Majorana neutrino masses
violate lepton number conservation, and in the standard model,
assuming only Higgs doublets are present, are
forbidden at the renormalizable level by gauge invariance.
The idea of the simplest version of the see-saw mechanism is to assume
that such terms are zero to begin with, but are generated effectively,
after right-handed neutrinos are introduced \cite{seesaw}.

If we introduce right-handed neutrino fields then there are two sorts
of additional neutrino mass terms that are possible. There are
additional Majorana masses of the form
\beq
M_\mathrm{RR}\overline{\nu_\mathrm{R}}\nu_\mathrm{R}^\ChargeC\; ,
\label{MRR}
\eeq
where $\nu_\mathrm{R}$ is a right-handed neutrino field and $\nu_\mathrm{R}^\ChargeC$ is
the CP conjugate of a right-handed neutrino field, in other words
a left-handed antineutrino field. In addition there are
Dirac masses of the form
\beq
m^\nu_\mathrm{LR}\overline{\nu_\mathrm{L}}\nu_\mathrm{R}\;.
\label{mLR}
\eeq
Such Dirac mass terms conserve lepton number, and are not forbidden
by electric charge conservation even for the charged leptons and
quarks.

Once this is done then the types of neutrino mass discussed
in Eq.\ref{MRR}, \ref{mLR} (but not Eq.\ref{mLL} since we do not 
assume direct mass terms, e.g.\ from Higgs triplets, at this stage) 
are permitted, and we have the mass matrix
\begin{equation}
\left(\begin{array}{cc} \overline{\nu_\mathrm{L}} & \overline{\nu^\ChargeC_\mathrm{R}}
\end{array} \\ \right)
\left(\begin{array}{cc}
0 & m^\nu_\mathrm{LR}\\
m^{\nu T}_\mathrm{LR} & M_\mathrm{RR} \\
\end{array}\right)
\left(\begin{array}{c} \nu_\mathrm{L}^\ChargeC \\ \nu_\mathrm{R} \end{array} \\
\right).
\label{matrix}
\end{equation}
Since the right-handed neutrinos are electroweak singlets
the Majorana masses of the right-handed neutrinos $M_\mathrm{RR}$
may be orders of magnitude larger than the electroweak
scale. In the approximation that $M_\mathrm{RR}\gg m^\nu_\mathrm{LR}$
the matrix in Eq.\ref{matrix} may be diagonalized to
yield effective Majorana masses of the type in Eq.\ref{mLL},
\beq
m^\nu_\mathrm{LL}=-m^\nu_\mathrm{LR}M_\mathrm{RR}^{-1}m^{\nu T}_\mathrm{LR}\; .
\label{seesaw}
\eeq
The effective left-handed Majorana masses $m^\nu_\mathrm{LL}$ are naturally
suppressed by the heavy scale $M_\mathrm{RR}$.
In a one family example, if we take $m^\nu_\mathrm{LR}=M_W$ and 
$M_\mathrm{RR}=M_{\mathrm{GUT}}$, 
then we find $m^\nu_\mathrm{LL}\sim 10^{-3}$ eV which looks good for solar
neutrinos.
Atmospheric neutrino masses would require
a right-handed neutrino with a mass below the GUT scale.

With three left-handed neutrinos and three right-handed neutrinos the
Dirac masses $m^\nu_\mathrm{LR}$ are a $3\times 3$ (complex) matrix
and the heavy Majorana masses $M_\mathrm{RR}$ form a separate $3\times
3$ (complex symmetric) matrix.  The light effective Majorana masses
$m^\nu_\mathrm{LL}$ are also a $3\times 3$ (complex symmetric) matrix
and continue to be given from Eq.\ref{seesaw} which is now interpreted
as a matrix product. From a model building perspective the fundamental
parameters which must be input into the see-saw mechanism are the
Dirac mass matrix $m^\nu_\mathrm{LR}$ and the heavy right-handed
neutrino Majorana mass matrix $M_\mathrm{RR}$.  The light effective
left-handed Majorana mass matrix $m^\nu_\mathrm{LL}$ arises as an
output according to the see-saw formula in Eq.\ref{seesaw}.

\begin{figure}
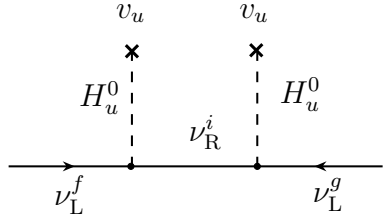

\begin{center} 
  \CenterEps[1]{typeIseesaw}\vphantom{\CenterEps[1]{typeIIseesaw}}
 \caption{\label{fig:TypeIDiagrams}
Diagram illustrating the type I see-saw mechanism.   
 } 
\end{center}
\end{figure}

The version of the see-saw mechanism discussed so far
is sometimes called the type I see-saw mechanism. It is the simplest
version of the see-saw mechanism, and can be thought of as resulting
from integrating out heavy right-handed neutrinos to produce
the effective dimension 5 neutrino mass
operator
\beq
-\frac{1}{4} (H_\mathrm{u} \cdot L^T) \,\kappa \, (H_\mathrm{u}\cdot L)\; ,
\label{dim5}
\eeq
where the dot indicates the $\SU (2)_\mathrm{L}$-invariant product 
and  
\beq
\kappa = 2 \,Y_{\nu} M_\mathrm{RR}^{-1} Y_{\nu}^T 
\label{kappa}
\eeq
with $Y_{\nu}$ being the neutrino Yukawa couplings and $m^\nu_\mathrm{LR}=Y_\nu
v_\mathrm{u}$ with $v_\mathrm{u} = \< H_\mathrm{u}\>$.   
The type I see-saw mechanism is illustrated diagramatically in 
Fig.~\ref{fig:TypeIDiagrams}.

\subsection{Type II See-Saw}
In models with a left-right 
 symmetric particle content like minimal left-right 
symmetric models, Pati-Salam models or grand unified theories (GUTs) 
based on $\SO (10)$, the type I see-saw mechanism is often  
generalized to a type II see-saw (see e.g.~\cite{Lazarides:1980nt}), 
where an additional direct 
mass term $m_{\mathrm{LL}}^{\mathrm{II}}$ for the light neutrinos is present. 

\begin{figure}
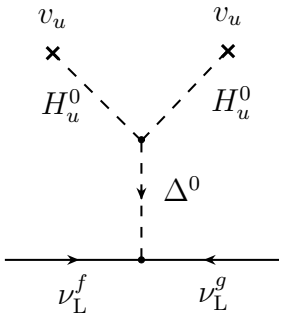

\begin{center} 
  \CenterEps[1]{typeIIseesaw}
 \caption{\label{fig:TypeIIDiagrams}
Diagram leading to a type II contribution $m^{\mathrm{II}}_{\mathrm{LL}}$ to 
the neutrino mass matrix via an induced vev of the neutral component of 
a triplet Higgs $\Delta$.  
 } 
\end{center}
\end{figure}

With such an additional direct mass term, 
the general neutrino mass matrix is given by
\begin{eqnarray}
\,
\left( \begin{array}{cc} \overline{\nu_{\mathrm{L}}}  &
  \overline{\nu^{\ChargeC }_\mathrm{R}}   \end{array}   \right) 
\left( \begin{array}{cc} 
m^{\mathrm{II}}_{\mathrm{LL}} \vphantom{\nu_{\mathrm{L}}^{\ChargeC }}&
m^\nu_{\mathrm{LR}}\\[1mm]
 m^{\nu T}_{\mathrm{LR}} \vphantom{\nu_{\mathrm{L}}^{\ChargeC }}&
 M_{\mathrm{RR}}
    \end{array}   \right)
  \,
 \left( \begin{array}{c} 
\nu_{\mathrm{L}}^{\ChargeC }  \\[1mm]
  {\nu_\mathrm{R}}   \end{array}   \right) .
\end{eqnarray}
Under the assumption that the mass eigenvalues $M_{\mathrm{R}i}$ of
$M_{\mathrm{RR}}$ are very large compared to the components of  $m^{\mathrm{II}}_{\mathrm{LL}}$
and $m_{\mathrm{LR}}$, the mass matrix can approximately be 
diagonalized yielding effective Majorana masses  
\begin{eqnarray}\label{eq:TypIIMassMatrix}
m^\nu_{\mathrm{LL}} \approx 
m^{\mathrm{II}}_{\mathrm{LL}} + m^{\mathrm{I}}_{\mathrm{LL}}
\end{eqnarray} 
with 
\begin{eqnarray}
m^{\mathrm{I}}_{\mathrm{LL}} \approx - m^\nu_{\mathrm{LR}}
\,M^{-1}_{\mathrm{RR}}\,m^{\nu T}_{\mathrm{LR}}
\end{eqnarray}
for the light neutrinos. 
The direct mass term  
$m^{\mathrm{II}}_{\mathrm{LL}}$ can also provide a naturally small contribution to the 
light neutrino masses if it stems e.g.~from a see-saw suppressed induced vev. 
We will refer to the general case,   
where both possibilities are allowed, as the II see-saw mechanism.
Realizing the type II contribution by generating  
the dimension 5 operator in Eq.\ref{dim5} via the exchange of heavy Higgs
triplets of SU(2)$_\mathrm{L}$ is illustrated diagrammatically in 
Fig.~\ref{fig:TypeIIDiagrams}.

\section{Sequential Right-Handed Neutrino Dominance in the Type I
  See-Saw Mechanism}

In this section we discuss an elegant and natural way of
accounting for a neutrino mass hierarchy and two large mixing angles,
called sequential dominance. The idea of sequential dominance is that one of
the right-handed neutrinos contributes dominantly to the
see-saw mechanism and determines the atmospheric neutrino
mass and mixing. A second right-handed neutrino
contributes sub-dominantly and determines the solar neutrino
mass and mixing. The third right-handed neutrino
is effectively decoupled from the see-saw mechanism.

\subsection{Single Right-Handed Neutrino Dominance }

Consider the case of full neutrino mass hierarchy
$m_3\gg m_2\gg m_1 \approx 0$.  
From Appendix~\ref{conventions} we see that in the diagonal
charged lepton basis, ignoring phases, the neutrino mass matrix 
is given by:
\begin{equation}
m^{\nu}_{LL}\approx
\left( \begin{array}{ccc}
m_2s_{12}^2 &
\frac{1}{\sqrt{2}}(m_2s_{12}c_{12}+m_3\theta_{13}) &
-\frac{1}{\sqrt{2}}(m_2s_{12}c_{12}-m_3\theta_{13}) \\
. &
\frac{1}{{2}}(m_3+ m_2c^2_{12})&
\frac{1}{{2}}(m_3- m_2c^2_{12}) \\
. & . & \frac{1}{{2}}(m_3+ m_2c^2_{12})
\end{array}
\right),
\label{approxmLL}
\end{equation}
neglecting terms like $m_2\theta_{13}$ and setting 
$\theta_{23}\approx \pi /4$.
Clearly this expression reduces to 
\begin{equation}
m^{\nu}_{LL}\approx
\left( \begin{array}{ccc}
0 & 0 & 0 \\
0 & 1 & 1 \\
0 & 1 & 1
\end{array}
\right)\frac{m}{2}\; ,
\label{1111}
\end{equation}
with $m=m_3$
in the approximation that $m_2$ and $\theta_{13}$ are
neglected. However the more exact
expression in Eq.\ref{approxmLL} shows that the required form of
$m_{LL}$ should have a very definite detailed structure.
The requirement $m_2 \ll m_3$ implies that
the sub-determinant of the mass matrix $m^{\nu}_{LL}$ is small:
\beq
\mbox{det}\, \left(
\begin{array}{cc}
m_{22} & m_{23}\\
m_{23} & m_{33} \\
\end{array}
\right)\ll m_3^2\;.
\label{det}
\eeq
This requirement in Eq.\ref{det}
is satisfied by Eq.\ref{approxmLL}, as may be readily
seen, and this condition must be reproduced in a natural way (without
fine-tuning) by any successful theory.

The goal of see-saw model building for hierarchical neutrino masses
is therefore to choose input
see-saw matrices $m^\nu_\mathrm{LR}$ and $M_\mathrm{RR}$ that will
give rise to the form in Eq.\ref{approxmLL}.
We now show how the input see-saw matrices can be simply chosen to
give this form, with the property of a
naturally small sub-determinant in Eq.\ref{det} using a mechanism
first suggested in 
\cite{King:1998jw}.\footnote{See also \cite{Davidson:1998bi}}
The idea was developed in \cite{King:1999cm} where it was called
single right-handed neutrino dominance (SRHND) . SRHND was first successfully
applied to the LMA MSW solution in \cite{King:1999mb}.


To understand the basic idea of dominance, it is 
instructive to begin by discussing
a simple $2\times 2$ example, where we have in mind applying
this to the atmospheric mixing in the 23 sector:
\begin{equation}
M_\mathrm{RR}=
\left( \begin{array}{cc}
Y & 0     \\
0 & X 
\end{array}
\right), \ \ \ \ 
m^\nu_\mathrm{LR}=
\left( \begin{array}{cc}
e & b \\
f & c
\end{array}
\right).
\label{2by2dom}
\end{equation}
The see-saw formula in Eq.\ref{seesaw} 
$m^\nu_\mathrm{LL}=-m^\nu_\mathrm{LR}M_\mathrm{RR}^{-1}m^{\nu T}_\mathrm{LR}$ 
gives 
\beq
-m^\nu_\mathrm{LL}
=
\left( \begin{array}{cc}
\frac{e^2}{Y}+\frac{b^2}{X}
& \frac{ef}{Y}+\frac{bc}{X} \\
\frac{ef}{Y}+\frac{bc}{X}
& \frac{f^2}{Y}+\frac{c^2}{X}
\end{array}
\right)
\approx
\left( \begin{array}{cc}
\frac{e^2}{Y}
& \frac{ef}{Y} \\
\frac{ef}{Y}
& \frac{f^2}{Y}
\end{array}
\right),
\label{one}
\eeq
where the approximation in Eq.\ref{one} 
assumes that the right-handed neutrino of mass $Y$
is sufficiently light that it dominates in the see-saw mechanism:
\beq
\frac{e^2,f^2,ef}{Y}\gg
\frac{b^2,c^2,bc}{X}\;.
\label{srhndp}
\eeq
The neutrino mass
spectrum from Eq.\ref{one}
then consists of one neutrino with mass $m_3\approx (e^2+f^2)/Y$
and one naturally light neutrino $m_2\ll m_3$, 
since the determinant of Eq.\ref{one} is clearly
approximately vanishing, due to the dominance assumption
\cite{King:1998jw}. 
The atmospheric angle from Eq.\ref{one} is
$\tan \theta_{23} \approx e/f$ \cite{King:1998jw}
which can be large or maximal providing $e \approx f$,
even in the case $e,f,b\ll c$ that 
the neutrino Dirac mixing angles 
arising from Eq.\ref{2by2dom} are small.
Thus two crucial features, namely a neutrino mass hierarchy
$m_3^2\gg m_2^2$ and a 
large neutrino mixing angle $\tan \theta_{23} \approx 1$, 
can arise naturally from the 
see-saw mechanism assuming the dominance of a single right-handed neutrino.
It was also realized that small perturbations from
the sub-dominant right-handed neutrinos can then lead to a small
solar neutrino mass splitting \cite{King:1998jw}, as we now discuss.

\subsection{Sequential Right-Handed Neutrino Dominance}\label{sec:SRHND}

In order to account for the solar and other mixing angles,
we must generalize the above discussion to the
$3\times 3$ case.
The SRHND mechanism is most simply described
assuming three right-handed neutrinos
in the basis where the right-handed neutrino mass matrix is diagonal
although it can also be developed in other bases
\cite{King:1999cm,King:1999mb}. In this basis we write the input
see-saw matrices as
\begin{equation}
M_\mathrm{RR}=
\left( \begin{array}{ccc}
X & 0 & 0    \\
0 & Y & 0 \\
0 & 0 & Z
\end{array}
\right)
\label{seq1}
\end{equation}
\begin{equation}
m^\nu_\mathrm{LR}=
\left( \begin{array}{ccc}
a & d & p    \\
b & e & q\\
c & f & r
\end{array}
\right)
\label{dirac}
\end{equation}
Each right-handed neutrino in the basis of Eq.\ref{seq1}
couples to a particular column of $m^\nu_\mathrm{LR}$ in Eq.\ref{dirac}.
There is no mass ordering of $X,Y,Z$ implied in Eq.\ref{seq1}.
In \cite{King:1998jw} it was suggested that one of the right-handed neutrinos
may dominate the contribution to $m^\nu_\mathrm{LL}$ if it is lighter than
the other right-handed neutrinos.
The dominance condition was subsequently generalized to
include other cases where the right-handed neutrino may be
heavier than the other right-handed neutrinos but dominates due to its larger
Dirac mass couplings \cite{King:1999cm}.
In any case the dominant right-handed neutrino may be taken to be the one
with mass $Y$ without loss of generality.

It was subsequently shown how to
account for the LMA MSW solution with a large solar angle
\cite{King:1999mb} by careful consideration of the sub-dominant contributions.
Sequential dominance occurs when the
right-handed neutrinos dominate sequentially \cite{King:1999mb},
\beq \label{eq:SeqRHNuDomCond}
\frac{|e^2|,|f^2|,|ef|}{Y}\gg
\frac{|xy|}{X} \gg
\frac{|x'y'|}{Z}\; ,
\label{srhnd}
\eeq
which is the straightforward generalization of Eq.\ref{srhndp}
where $x,y\in a,b,c$ and $x',y'\in p,q,r$.
Assuming SRHND with sequential sub-dominance as in
Eq.\ref{srhnd}, then Eq.\ref{seesaw}, \ref{seq1},
\ref{dirac} give
\beq
-m^\nu_\mathrm{LL}
\approx
\left( \begin{array}{ccc}
\frac{a^2}{X}+\frac{d^2}{Y}
& \frac{ab}{X}+ \frac{de}{Y}
& \frac{ac}{X}+\frac{df}{Y}    \\
.
& \frac{b^2}{X}+\frac{e^2}{Y}
& \frac{bc}{X}+\frac{ef}{Y}    \\
.
& .
& \frac{c^2}{X}+\frac{f^2}{Y}
\end{array}
\right),
\label{mLL2}
\eeq
where the contribution from the right-handed neutrino of mass $Z$ may be
neglected according to Eq.\ref{srhnd}.
If the couplings satisfy the sequential dominance
condition in Eq.\ref{srhnd}
then the matrix in Eq.\ref{mLL2} resembles the Type IA matrix,
and furthermore has a naturally small sub-determinant as in
Eq.\ref{det}. 
This leads to a full neutrino mass hierarchy
\beq
m_3^2\gg m_2^2\gg m_1^2
\eeq
and, ignoring phases, the solar angle only depends
on the sub-dominant couplings and is given by
$\tan \theta_{12} \approx a/(c_{23}b-s_{23}c)$ \cite{King:1999mb}.
The simple requirement for large solar angle is then $a\sim b-c$
\cite{King:1999mb}.

Including phases the neutrino masses
are given to leading order in $m_2/m_3$ by diagonalizing
the mass matrix in Eq.\ref{mLL2} using the analytic procedure
described in Appendix D of \cite{King:2002nf}.
In the case that $d=0$, corresponding to a 11 texture zero
in Eq.\ref{dirac}, we have \cite{King:2002nf,King:2002qh}:
\bea
m_1 & \sim & O(\frac{x'y'}{Z}) \label{m1} \; ,\\
m_2 & \approx &  \frac{|a|^2}{Xs_{12}^2} \label{m2} \; ,\\
m_3 & \approx & \frac{|e|^2+|f|^2}{Y}\; ,
\label{m3}
\eea
where $s_{12}=\sin \theta_{12}$ is given below.
Note that with SD each neutrino mass is generated
by a separate right-handed neutrino, and the sequential dominance condition
naturally results in a neutrino mass hierarchy $m_1\ll m_2\ll m_3$.
The neutrino mixing angles are given to leading order in $m_2/m_3$ by 
\cite{King:2002nf,King:2002qh}:
\bea
\label{eq:SRND_t23}\tan \theta_{23} & \approx & \frac{|e|}{|f|} \label{23}\; ,\\
\label{eq:SRND_t12}\tan \theta_{12} & \approx &
\frac{|a|}
{c_{23}|b|
\cos(\tilde{\phi}_b)-
s_{23}|c|
\cos(\tilde{\phi}_c)} \label{12} \; ,\\
\label{eq:SRND_t13}\theta_{13} & \approx &
e^{i(\tilde{\phi}+\phi_a-\phi_e)}
\frac{|a|(e^*b+f^*c)}{[|e|^2+|f|^2]^{3/2}}
\frac{Y}{X}\; ,
\label{13}
\eea
where we have written some (but not all) complex Yukawa couplings as
$x=|x|e^{i\phi_x}$. The phase $\delta$
is fixed to give a real angle
$\theta_{12}$ by,
\beq
c_{23}|b|
\sin(\tilde{\phi}_b)
\approx
s_{23}|c|
\sin(\tilde{\phi}_c)\; ,
\label{chi1}
\eeq
where
\bea
\tilde{\phi}_b &\equiv &
\phi_b-\phi_a-\tilde{\phi}+\delta, \nonumber \; ,\\
\tilde{\phi}_c &\equiv &
\phi_c-\phi_a+\phi_e-\phi_f-\tilde{\phi}+\delta\; .
\label{bpcp}
\eea
The phase $\tilde{\phi}$
is fixed to give a real angle
$\theta_{13}$ by 
\beq
\tilde{\phi} \approx  \phi_e-\phi_a -\arg(e^*b+f^*c)\; .
\label{phi2dsmall}
\eeq
Physically these results show that in sequential dominance 
the atmospheric neutrino mass $m_3$ and mixing $\theta_{23}$
is determined by the
couplings of the dominant right-handed neutrino of mass $Y$.
The solar neutrino mass $m_2$ and mixing $\theta_{12}$
is determined by the couplings of the
sub-dominant right-handed neutrino of mass $X$.
The third right-handed neutrino of mass $Z$
is effectively decoupled from the see-saw mechanism
and leads to the vanishingly small mass $m_1\approx 0$.

\subsection{Types of Sequential Right-Handed Neutrino Dominance}

Assuming sequential dominance,
there is still an ambiguity regarding the mass ordering of the
heavy Majorana right-handed neutrinos.
So far we have assumed that the dominant right-handed neutrino
of mass $Y$ is dominant because it is the lightest one.
We emphasize that this need not be the case.
The neutrino of mass $Y$ could be dominant even if it is the heaviest
right-handed neutrino, providing its Yukawa couplings are strong
enough to overcome its heaviness and satisfy the condition in
Eq.\ref{srhnd}. In hierarchical mass matrix models, it is
natural to order the right-handed neutrinos so that the heaviest
right-handed neutrino is the third one, the intermediate right-handed
neutrino is the second one, and the lightest right-handed neutrino
is the first one. It is also natural to assume that the 33 Yukawa
coupling is of order unity, due to the large top quark mass.
It is therefore possible that the dominant right-handed
neutrino is the heaviest (called heavy sequential dominance or
HSD), the lightest (called light sequential dominance or LSD), or
the intermediate one (called intermediate sequential dominance or
ISD). This leads to the six possible types of sequential dominance
corresponding to the six possible mass orderings of the right-handed
neutrinos as shown in Table~\ref{table1}.
In each case the dominant right-handed neutrino is the one with mass
$Y$, and the leading sub-dominant right-handed neutrino is the one
with mass $X$. The resulting see-saw matrix $m^\nu_\mathrm{LL}$ is invariant
under re-orderings of the right-handed neutrino columns,
but the leading order form of the neutrino Yukawa matrix $Y_{\nu}$
is not.

\begin{table}[ht]
\begin{center}
\begin{eqnarray*}\hspace{1.00cm}
\begin{array}{|c|c|c|c|}
\hline
\hline
\mbox{Type of SD} & M_\mathrm{RR}  & m^\nu_\mathrm{LR}=Y_{\nu} v_\mathrm{u} 
& \mbox{Leading $Y_{\nu}$}\\
\hline
\hline
 \stackrel{\rm LSDa}{Y<X<Z}   &
 \left(
\begin{array}{ccc}
Y & 0 & 0 \\
0 & X & 0\\
0 & 0 & Z \\
\end{array}
\right) 
&
 \left(
\begin{array}{ccc}
d & a & p\\
e & b & q\\
f & c & r\\
\end{array}
\right) 
&
 \left(
\begin{array}{ccc}
0 & 0 & 0 \\
0 & 0 & 0\\
0 & 0 & 1 \\
\end{array}
\right) 
\\
\hline
\hline
 \stackrel{\rm LSDb}{Y<Z<X}   &
 \left(
\begin{array}{ccc}
Y & 0 & 0 \\
0 & Z & 0\\
0 & 0 & X \\
\end{array}
\right) 
&
 \left(
\begin{array}{ccc}
d & p & a \\
e & q & b\\
f & r & c \\
\end{array}
\right) 
&
 \left(
\begin{array}{ccc}
0 & 0 & 1\\
0 & 0 & 1\\
0 & 0 & 1 \\
\end{array}
\right) 
\\
\hline
\hline
 \stackrel{\rm ISDa}{X<Y<Z}   &
 \left(
\begin{array}{ccc}
X & 0 & 0 \\
0 & Y & 0\\
0 & 0 & Z \\
\end{array}
\right) 
&
 \left(
\begin{array}{ccc}
a & d & p \\
b & e & q \\
c & f & r \\
\end{array}
\right) 
&
 \left(
\begin{array}{ccc}
0 & 0 & 0 \\
0 & 0 & 0\\
0 & 0 & 1 \\
\end{array}
\right) 
\\
\hline
\hline
 \stackrel{\rm ISDb}{Z<Y<X}   &
 \left(
\begin{array}{ccc}
Z & 0 & 0 \\
0 & Y & 0\\
0 & 0 & X \\
\end{array}
\right) 
&
 \left(
\begin{array}{ccc}
p & d & a \\
q & e & b\\
r & f & c \\
\end{array}
\right) 
&
 \left(
\begin{array}{ccc}
0 & 0 & 1 \\
0 & 0 & 1\\
0 & 0 & 1 \\
\end{array}
\right) 
\\
\hline
\hline
 \stackrel{\rm HSDa}{Z<X<Y}   &
 \left(
\begin{array}{ccc}
Z & 0 & 0 \\
0 & X & 0\\
0 & 0 & Y \\
\end{array}
\right) 
&
 \left(
\begin{array}{ccc}
p & a & d \\
q & b & e\\
r & c & f \\
\end{array}
\right) 
&
 \left(
\begin{array}{ccc}
0 & 0 & 0 \\
0 & 0 & 1\\
0 & 0 & 1 \\
\end{array}
\right) 
\\
\hline
\hline
 \stackrel{\rm HSDb}{X<Z<Y}   &
 \left(
\begin{array}{ccc}
X & 0 & 0 \\
0 & Z & 0\\
0 & 0 & Y \\
\end{array}
\right) 
&
 \left(
\begin{array}{ccc}
a & p & d \\
b & q & e\\
c & r & f \\
\end{array}
\right) 
&
 \left(
\begin{array}{ccc}
0 & 0 & 0 \\
0 & 0 & 1\\
0 & 0 & 1 \\
\end{array}
\right) \\
\hline
\hline
\end{array}
\end{eqnarray*}
\end{center}
\caption{\label{table1}
Types of sequential dominance (SD), classified
according to the mass ordering of the right-handed neutrinos.
Light sequential dominance (LSD) corresponds to the dominant
right-handed neutrino of mass $Y$ being the lightest.
Intermediate sequential dominance (ISD) corresponds to the dominant
right-handed neutrino of mass $Y$ being the intermediate one.
Heavy sequential dominance (HSD) corresponds to the dominant
right-handed neutrino of mass $Y$ being the heaviest. 
The fourth column of the table shows the leading order form for 
$Y_\nu$ under the assumption of a large 33-element in the Yukawa matrix.  
}
\end{table}	

It is worth emphasizing that since all the forms above give the
same light effective see-saw neutrino matrix $m^\nu_\mathrm{LL}$ in
Eq.\ref{mLL2}, under the sequential dominance assumption
in Eq.\ref{srhnd}, this implies that the analytic results for neutrino masses
and mixing angles applies to all of these forms.
They are distinguished theoretically by different preferred leading order
forms of the neutrino Yukawa matrix $Y_{\nu}$ shown in the table.
These leading order forms follow from the
the large mixing angle requirements $e\sim f$ and
$a \sim b-c$.\footnote{Note that the leading order $Y_{\nu}$ in Table~\ref{table1}
only gives the independent order unity entries in the matrix,
so that for example in LSDb we would expect $b-c\sim 1$ in general,
and not zero.}
Thus we see that LSDa, and ISDa are
consistent with a form of Yukawa matrix with small Dirac mixing
angles, while HSDa and HSDb correspond to the so called ``lop-sided''
forms.

\section{Sequential Right-Handed Lepton Dominance
 in the Type I See-Saw Mechanism}\label{sec:SmallNuMixing}

In this section we show how bi-large mixing could originate from
the charged lepton sector using a 
generalization of sequential right-handed neutrino dominance 
\cite{King:1999mb,King:2002nf} to all
right-handed leptons \cite{Antusch:2004re}. 
We write the mass matrices for the charged leptons  
$m_\mathrm{E}$ as 
\begin{eqnarray}
m_\mathrm{E} = \left(
\begin{array}{ccc}
p' & d' & a' \\
q' & e' & b' \\
r' & f' & c'
\end{array}
\right)\!.
\end{eqnarray}
In our notation, 
each right-handed charged lepton couples to a column in
$m_\mathrm{E}$.
For the charged leptons, the sequential dominance conditions are
\cite{Antusch:2004re}:
\begin{eqnarray}\label{eq:SeqDominanceCL}
|a'|,|b'|,|c'| \gg |d'|,|e'|,|f'| \gg |p'|,|q'|,|r'| \;.
\end{eqnarray}
They imply the desired hierarchy for the charged lepton masses $m_\tau
\gg m_\mu \gg m_e$ and small right-handed mixing of
$U_{e_\mathrm{R}}$. 
We assume zero mixing from the neutrino sector which corresponds to 
the MNS matrix being given by $U_\mathrm{MNS}=U_{e_{\mathrm{L}}}\!\!\cdot
\mbox{diag}\,(1,e^{i \beta^{\nu}_2},e^{i \beta^{\nu}_3})$
in the conventions in Appendix \ref{conventions}.
A natural possibility 
for obtaining a small $\theta_{13}$ is \cite{Antusch:2004re}
\begin{eqnarray}\label{eq:CondForSmallT13}
|d'|,|e'| \ll |f'| \; .
\end{eqnarray} 
In leading order in $|d'|/|f'|$ and $|e'|/|f'|$, 
for the mixing angles $\theta_{12},\theta_{23}$ and $\theta_{13}$, we obtain   
\begin{subequations}\label{eq:mixings_1b}\begin{eqnarray}
\label{eq:t12_C2}\tan (\theta_{12}) &\approx&  \frac{|a'|}{ |b'|} \; ,\\
\label{eq:t23_C2}\tan (\theta_{23}) &\approx&   \frac{  s_{12}\, |a'| +  c_{12}
\,|b'| }{
|c'| } \;,\vphantom{\frac{f}{f}}\\
\label{eq:t13_C2}\tan (\theta_{13}) &\approx&  \frac{
s_{12}\,|e'| \, e^{i (\phi_{a'} - \phi_{b'} + \phi_{e'} +\delta)} - 
c_{12}|d'| \, e^{i (\phi_{d'} + \delta)} 
}{
|f'|\,e^{i (\phi_{a'} - \phi_{c'} +  \phi_{f'})}}  \vphantom{\frac{f}{f}}\; ,
\end{eqnarray}\end{subequations}
where the Dirac CP phase $\delta$ is determined such that $\theta_{13}$ 
is real, which requires 
\begin{eqnarray}\label{eq:mixings_1c}
\label{eq:delta_C2}\tan (\delta) \approx 
\frac{c_{12}\, |d'| \,\sin (\phi_{a'} - \phi_{c'} - \phi_{d'} + \phi_{f'}) - 
s_{12}\, |e'|\,
\sin (\phi_{b'} - \phi_{c'} - \phi_{e'} + \phi_{f'})}{
c_{12} \,|d'| \,\cos (\phi_{a'} - \phi_{c'} - \phi_{d'} + \phi_{f'}) - 
s_{12}\, |e'| \,\cos
(\phi_{b'} - \phi_{c'} - \phi_{e'} + \phi_{f'})} 
\, .\vphantom{\frac{f}{f}}
\end{eqnarray}
Given $\tan (\delta)$, $\delta$ has to be chosen such that 
$\tan(\theta_{13})\ge 0$ in order to match with the usual convention
$\theta_{13}\ge 0$. 
The phases $\beta^e_2$ and $\beta^e_3$ from the charge lepton
sector are given by 
 \begin{subequations}\label{eq:mixings_1a}\begin{eqnarray}
\label{eq:b2_C2}\beta^e_2 &\approx& \phi_{a'} - \phi_{b'} + \pi \; , \\
\label{eq:b3_C2}\beta^e_3 &\approx& \phi_{a'} - \phi_{c'} \; . 
\end{eqnarray}\end{subequations}
Note that in the case that the neutrino sector induces Majorana phases, the 
total Majorana phases $\beta_2$ and $\beta_3$ of the MNS matrix are 
given by 
 \begin{subequations}\label{eq:mixings_1d}\begin{eqnarray}
\label{eq:total_b2_C2}\beta_2 &\approx& \beta^e_2 + \beta^\nu_2 \; ,\\
\label{eq:total_b3_C2}\beta_3 &\approx& \beta^e_3 + \beta^\nu_3 \; .
\end{eqnarray}\end{subequations} 
$\theta_{13}$ only depends on $d'/f'$ and $e'/f'$ from 
the Yukawa couplings to the sub-dominant right-handed muon and on $\theta_{12}$.  
We find that in the limit $|d'|,|e'| \ll |f'|$, the two large mixing angles 
$\theta_{12}$ and $\theta_{23}$ approximately depend only on 
$a'/c'$ and $b'/c'$ from 
the right-handed tau Yukawa couplings.  
Both mixing angles are large if $a',b'$ and $c'$ are of the same order. 

In addition to achieving bi-large mixing from the charged lepton
sector, we also require now small mixing from the neutrino sector.
Usually, sequential RHND \cite{King:1999mb,King:2002nf} is viewed as a
framework for generating large solar mixing $\theta_{12}$ and large
atmospheric mixing $\theta_{23}$ in the neutrino mass matrix.
However, given sequential dominance in the neutrino sector in
Eq.\ref{eq:SeqRHNuDomCond} which guarantees a neutrino mass hierarchy,
one can
easily find the conditions for small mixing from the neutrinos as well
from Eq.\ref{eq:SRND_t23},\ref{eq:SRND_t12}. Using the notation of
subsection \ref{sec:SRHND}, we need $d,e \ll f$ and $a\ll b,c$.
Small mixing from the neutrino sector thus requires three small entries in 
$m^\nu_\mathrm{LR}$. 
As shown in \cite{Antusch:2004xd},  
three zero entries in $Y_\nu$ might stem from a 
spontaneously broken SO(3) flavour symmetry and real vacuum alignment. 
Other realizations might by found via Abelian or discrete symmetries.

\section{Type II See-Saw Upgrade}
In type I see-saw models, it seems to be difficult to obtain a
partially degenerate or quasi-degenerate neutrino mass spectrum in a
natural way, whereas hierarchical masses seem to be natural.  The
direct mass term in type II models on the other hand has the potential
to provide a natural way for generating neutrino masses with a partial
degeneracy. In this section we show that it is possible to obtain 
a partially degenerate neutrino mass spectrum by essentially adding
a type II direct neutrino mass 
contribution proportional to the unit matrix. In this case the
neutrino mass scale is controlled by the type II direct mass term,
while the neutrino mass splittings 
(which are generally now much smaller) and mixings continue to be
determined by the type I see-saw matrix using sequential dominance
as described earlier.

Thus we shall consider a type II upgrade 
\cite{Antusch:2004xd}, where the mass matrix of the light neutrinos 
in Eq.\ref{eq:TypIIMassMatrix} has the particular form    
\begin{eqnarray}
m^\nu_{\mathrm{LL}} \approx m^{\mathrm{II}}\, 
\left(\begin{array}{ccc}
1&0&0\\
0&1&0\\
0&0&1
\end{array}\right)
 +
 \left(\begin{array}{ccc}
(m^{\mathrm{I}}_{\mathrm{LL}})_{11} &(m^{\mathrm{I}}_{\mathrm{LL}})_{12} & (m^{\mathrm{I}}_{\mathrm{LL}})_{13}\\
(m^{\mathrm{I}}_{\mathrm{LL}})_{21} &(m^{\mathrm{I}}_{\mathrm{LL}})_{22} &(m^{\mathrm{I}}_{\mathrm{LL}})_{23}\\
(m^{\mathrm{I}}_{\mathrm{LL}})_{31} &(m^{\mathrm{I}}_{\mathrm{LL}})_{32} &(m^{\mathrm{I}}_{\mathrm{LL}})_{33}
\end{array}\right),
\end{eqnarray}  
in the basis where the mass matrix $M_{\mathrm{RR}}$ of the heavy right-handed
neutrinos is diagonal.
  


 To understand the effect of the type II contribution
 $m_{\mathrm{LL}}^{\mathrm{II}}$, we consider the diagonalization of
 $m_{\mathrm{LL}}^{\mathrm{I}}$ by a unitary transformation
 $(m_{\mathrm{LL}}^{\mathrm{I}})_{\mathrm{diag}} = V
 m_{\mathrm{LL}}^{\mathrm{I}} V^T $.  If we assume for the moment that
 the type I see-saw mass matrix $m_{\mathrm{LL}}^{\mathrm{I}}$ is
 real, which implies that $V$ is an orthogonal matrix, we obtain
  \begin{eqnarray}\label{eq:TypIISeeSawFormulaUnitMatrix}
 (m^\nu_{\mathrm{LL}})_{\mathrm{diag}} \;= \; 
 m^{\mathrm{II}}\, V V^T +  V  m_{\mathrm{LL}}^{\mathrm{I}} V^T
\;=\; m^{\mathrm{II}}\, \mathbbm{1} + (m_{\mathrm{LL}}^{\mathrm{I}})_{\mathrm{diag}}
\; . 
 \end{eqnarray}
The additional direct mass term leaves the predictions for the mixings
from the type I see-saw contribution unchanged in this case.  This
allows to transform many type I see-saw models for hierarchical
neutrino masses into type II see-saw models for partially degenerate
or quasi-degenerate neutrino masses while maintaining the predictions
for the mixing angles.  Obviously, in the general complex case, it is
no longer that simple since for a unitary matrix $V V^T \not=
\mathbbm{1}$ and the phases will have impact on the predictions for
the mixings.  However, as we will see below, with sequential
right-handed neutrino dominance \cite{King:1999mb,King:2002nf} for the
type I contribution to the neutrino mass matrix, and a particular
phase structure, the known techniques and mechanisms for explaining
the bi-large lepton mixings can be directly applied also in the
presence of CP phases.

\subsection{Type II Upgrade of a ISD Model}

 As an example of a type II model where the bi-large lepton mixing stems 
from the neutrino mass matrix, we now consider explicitly the model A1 
of table 4 in \cite{Antusch:2004xd} with sequential right-handed neutrino 
dominance \cite{King:1999mb,King:2002nf} for the type 
I part $m_\nu^{\mathrm{I}}$ of the neutrino mass matrix. 
 The leading order Dirac mass matrices are  
\begin{eqnarray}\label{eq:CommonPhases}\label{eqn:YukawaMatrices_A1}
m^\nu_\mathrm{LR}
= 
\left(\begin{array}{ccc}
a\,e^{i \da}&0&0\\
b\,e^{i \da}&e\,e^{i \db}&0\\
c\,e^{i \da}&f\,e^{i \db}&r\,e^{i \dc}
\end{array}
\right)
   \!, \;\;
 m_\mathrm{E} 
= 
\left(\begin{array}{ccc}
a'\,e^{i \daP}&0&0\\
b'\,e^{i \daP}&e'\,e^{i \dbP}&0\\
c'\,e^{i \daP}&f'\,e^{i \dbP}&r'\,e^{i \dcP}
\end{array}
\right)
 \!,
\end{eqnarray}
where here $a,b,c,e,f,r$ and $a',b',c',e',f',r'$ are real.
$M_\mathrm{RR}$ and  the type II contribution  
$m^{\mathrm{II}}_{\mathrm{LL}}$ are given by 
\begin{eqnarray}
M_\mathrm{RR}
=:
\left(\begin{array}{ccc}
\!X&0&0\!\\
\!0&Y&0\!\\
\!0&0&Z\!
\end{array}
\right)\!,\;\;
m^{\mathrm{II}}_{\mathrm{LL}} = m^{\mathrm{II}}\, 
\left(\begin{array}{ccc}
1&0&0\\
0&1&0\\
0&0&1
\end{array}\right),
\end{eqnarray}
denoting the mass of the dominant right-handed neutrino by $Y$ and the mass of
the sub-dominant one by $X$.  
The sequential RHND condition we impose is then 
\begin{eqnarray}\label{eq:SequSubDominace}
\left|\frac{e^2,f^2}{Y}\right| \gg 
\left|\frac{a^2,b^2,c^2}{X}\right|\gg 
\left|\frac{r^2}{Z}\right| \; .
\end{eqnarray}
The leading order type II neutrino mass matrix is given by 
the type II see-saw formula of Eq.\ref{eq:TypIIMassMatrix}.  

 The masses of the charged leptons are given by 
 $m_\tau = r',m_\mu = e'$  and 
 $m_e = a'$.
In addition we note that 
 the mixings $\theta^e_{12},\theta^e_{13}$ and $\theta^e_{23}$, which stem from 
 $U_{e_\mathrm{L}}$ and could
 contribute to the MNS matrix, are very small. 
 Furthermore, in leading order each column of $M_e$ has a common complex phase, which can be 
 absorbed by $U_{e_\mathrm{R}}$. Therefore, the
 charged leptons do not influence the leptonic CP phases in this approximation.

Using the analytical methods for diagonalizing neutrino mass matrices with small $\theta_{13}$
derived in \cite{King:2002nf}, from 
$m^\nu_{\mathrm{LL}} = m_{\mathrm{LL}}^{\mathrm{II}} + 
m_{\mathrm{LL}}^{\mathrm{I}}$ we find for the mixing 
angles
 \begin{subequations}\begin{eqnarray}
\label{AnalyticResultForT23} \tan (\theta_{23}) &\approx& \frac{|e|}{|f|} \; , \\
\label{AnalyticResultForT13} \tan (2 \theta_{13}) &\approx& 
\frac{2 \,|a|  }{X} \,
\frac{|\sin (\theta_{23}) |b| + \cos (\theta_{23}) |c| \,\mbox{sign}\,(b\, c\,
e\, f)|}{
|2\,  m^{\mathrm{II}}\, \sin (\widetilde \delta) + m_3^{\mathrm{I}}
e^{i(2\db + 3\pi/2  -\widetilde \delta)}|
}  \; ,\\
\label{AnalyticResultForT12}\tan (\theta_{12}) &\approx& \frac{|a|}{|\cos (\theta_{23}) |b| - 
\sin (\theta_{23}) |c|\,\mbox{sign}\,(b\, c\,
e\, f)| } \; ,
\end{eqnarray}\end{subequations} 
where $m_i^{\mathrm{I}}$ ($i\in \{1,2,3\}$) are the mass eigenvalues of the hierarchical
$m_{\mathrm{LL}}^{\mathrm{I}}$ given by
 \begin{subequations}\begin{eqnarray}
m_1^{\mathrm{I}} &=&  
{\cal O} \left( \frac{r^2}{Z} \right)   \;\approx\;0 \; , \vphantom{\frac{|e|}{|f|}}\\
\label{eq:m2I_A1} m_2^{\mathrm{I}} &\approx& \frac{(|a|^2 + |\cos (\theta_{23}) |b| - 
\sin (\theta_{23}) |c|\,\mbox{sign}\,(b\, c\,
e\, f)|^2) }{X}
\;\approx\;\frac{|a|^2 }{\sin^2 (\theta_{12}) X}\;,\\
m_3^{\mathrm{I}} &\approx& \frac{(|e| \sin (\theta_{23})+|f|\cos (\theta_{23}))^2 }{Y}\;, 
\end{eqnarray}\end{subequations} 
and with $\widetilde \delta$ defined by 
\begin{eqnarray}\label{eq:DiracCP_A1_1}
 \tan (\widetilde \delta) := 
\frac{m_3^{\mathrm{I}} \, \sin (2 \db -  2 \da)}{  
m_3^{\mathrm{I}} \,\cos (2 \db -  2 \da) - 2 \, m^{\mathrm{II}} \, \cos (2 \da)}\, .
\end{eqnarray}
Given $\tan (\widetilde \delta)$, $\widetilde \delta$ has to be chosen such that 
\begin{eqnarray}
\frac{\sin (\theta_{23}) |b| + \cos (\theta_{23}) |c| \,\mbox{sign}\,(b\, c\,
e\, f)}{
\mbox{sign}\,(a\, b) \, [
2\,  m^{\mathrm{II}}\,e^{-i (2\da+ 3\pi/2)} \, \sin (\widetilde \delta) + m_3^{\mathrm{I}}
e^{-i(2\da -2\db+\widetilde \delta)}
]
}
\ge 0 \;.
\end{eqnarray}
This does not effect $\theta_{13}$, which we have defined to be $\ge 0$, however
it is relevant for extracting the Dirac CP phase $\delta$, 
given by 
\begin{eqnarray}\label{eq:DiracCP_A1_2}
\delta &\approx& 
\left\{
\begin{array}{cl}
\widetilde \delta & \mbox{for $P\ge0$\,,}\\
\widetilde \delta + \pi & \mbox{for $P<0$\,,}
\end{array}
\right.
\end{eqnarray} 
with $P$ being defined by
\begin{eqnarray}
P := \frac{
\cos (\theta_{23}) |b| - 
\sin (\theta_{23}) |c|\,\mbox{sign}\,(b\, c\,
e\, f)
}{
\mbox{sign}\,(a\, b) \, [
(\cos (\theta_{23}) |b| - 
\sin (\theta_{23}) |c|\,\mbox{sign}\,(b\, c\,
e\, f))^2 - |a|^2
]
}\; .
\end{eqnarray}
The mass eigenvalues of the complete type II neutrino mass matrix are given by
 \begin{subequations}\label{eq:ComplMassEigenvOfMnuTypeII}\begin{eqnarray}
m_1  &\approx&  |m^{\mathrm{II}}|\;,\\
m_2  &\approx& | m^{\mathrm{II}} - m_2^{\mathrm{I}} \,e^{i 2 \da}|\;,\\
m_3  &\approx& | m^{\mathrm{II}} - m_3^{\mathrm{I}} \,e^{i 2 \db}|\;, 
\end{eqnarray} \end{subequations} 
and, for $m^{\mathrm{II}} \not= 0$, the Majorana phases $\beta_2$ and $\beta_3$  
can be extracted by
\begin{subequations}\label{eq:MajPhases_A1}\begin{eqnarray}
\beta_2  &\approx& \frac{1}{2}\mbox{arg}\,( m^{\mathrm{II}} - m_2^{\mathrm{I}} \,e^{i 2 \da})\;,\\
\beta_3  &\approx& \frac{1}{2}\mbox{arg}\,( m^{\mathrm{II}} - m_3^{\mathrm{I}} \,e^{i 2 \db})\;.
\end{eqnarray}\end{subequations} 
In the classes of type II see-saw models with sequential right-handed 
neutrino dominance for the type I 
contribution to the neutrino mass matrix and real vacuum alignment leading to
the phase structure of the Yukawa matrices as in Eq.\ref{eq:CommonPhases}, 
the solar and the atmospheric neutrino mixings $\theta_{12}$ and $\theta_{23}$ 
are independent of the type II mass scale $m^{\mathrm{II}}$ and of the  
complex phases of the neutrino Yukawa matrix. We can thus 
upgrade these types of models continously from hierarchical neutrino 
mass spectra to 
partially degenerate ones, while maintaining the 
predictions for the two large lepton mixings. 

\section{Model Building Applications of Sequential \\ Dominance}

We have seen that sequential dominance is not a model, but
is a general sub-mechanism within the see-saw mechanism.
Sequential dominance may be used to obtain hierarchical type I
neutrino masses, together with bi-large lepton mixing,
in a completely natural way, overcoming the usual naturalness
objection to the see-saw mechanism in this case.
We have also seen that sequential dominance
may be extended to the case where the lepton mixing arises from
the charged lepton sector. Furthermore we have seen that 
the sequential dominance mechanism is also useful within the
framework of the type II see-saw mechanism in the case that the
additional type II mass contributions are proportional to the
unit matrix. Despite the successes of sequential dominance,
the conditions on which it is based have 
just been stated without any explanation. It also remains to 
be seen how the mechanism of sequential dominance can be
used to construct realistic unified models of flavour.
 
In this section we discuss some of the model building applications
of sequential dominance. We shall see that the use of sequential 
dominance is ideally suited to GUTs and family symmetry models,
has already been used in quite a number of works of this nature.
Sequential dominance also makes contact with studies based on
two right-handed neutrinos. Finally there have been some interesting
cosmological applications that have recently been proposed.
Given the simplicity and naturalness of sequential dominance,
it is reasonable to expect that it will continue to be exploited 
increasingly in the future.

\subsection{Effective Two Right-Handed Neutrino Models}
In sequential dominance we have seen that one of the right-handed
neutrinos effectively decouples from the see-saw mechanism.
Without loss of generality we have denoted the mass of this
decoupled right-handed neutrino as $Z$. From Table~\ref{table1}
we see that the decoupled right-handed neutrino of mass
$Z$ may be the lightest, the heaviest of the intermediate mass
right-handed neutrino. If it is the lightest or the
second lightest then it could in principle play an important
part in leptogenesis or inflation and so have cosmological relevance
even though it is decoupled from the see-saw mechanism.
However if it is the heaviest right-handed neutrino,
as in LSDa or ISDa in Table~\ref{table1},
then it would be expected to play no part in phenomenology.
In these cases, the heaviest neutrino of mass $Z$ is
completely decoupled from physics, and sequential
dominance reduces to effectively two right-handed neutrino models,
as pointed out in \cite{King:1999mb,King:2002qh}.
Recently there have been several studies based on the ``minimal see-saw'' 
involving two right-handed neutrinos \cite{Frampton:2002qc},
and it is worth bearing in mind that such models 
could naturally arise as the limiting case of sequential dominance.

\subsection{GUT and Family Symmetry Models}

There are many models in the literature based on single right-handed
neutrino dominance or sequential dominance.
For example explicit realisations of 
the small determinant condition of Altarelli and Feruglio
implicitly involve single right-handed neutrino dominance,
or sequential dominance,
together with U(1) family symmetry and
SU(5) GUTs \cite{Altarelli:1998nx}.
An example of sequential dominance of the HSD type
in Pati-Salam models with
U(1) family symmetry was considered in \cite{King:2000ge}.
Single right-handed neutrino dominance has also been applied to
SO(10) GUT models involving a U(2) family symmetry
\cite{Barbieri:1999pe}. Sequential dominance of the LSD type with
SU(3) family symmetry and SO(10) GUTs has been considered in
\cite{King:2001uz}. Type II up-gradable models
based on sequential dominance of the ISD type 
with $SO(3)$ family symmetry have been considered in 
\cite{Antusch:2004re,Antusch:2004xd}.
This list is not exhaustive, but represents a subset
of models based on single or sequential right-handed neutrino
dominance. The main point is that sequential dominance
can readily be included in a wide range GUT and family symmetry models,
and it enhances the naturalness of such models.

\subsection{Sneutrino Inflation Models}

Sequential dominance has recently also been applied to sneutrino
inflation \cite{Ellis:2003sq}. Requiring a low reheat temperature
after inflation, in order to solve the gravitino problem,
forces the sneutrino inflaton to couple very weakly to 
ordinary matter and its superpartner almost to decouple from the
see-saw mechanism.
This decoupling of a right-handed neutrino from the see-saw mechanism
is a characteristic of sequential dominance.

\section{Phenomenological Implications of Sequential \\ Dominance}
We now review phenomenological consequences of type I see-saw models with 
sequential dominance and their type II upgrades for the low energy 
neutrino parameters and high-energy mechanisms as leptogenesis and 
minimal lepton flavour violation (LFV).  
In order to compare the predictions of see-saw models based on sequential
dominance with the experimental data obtained at low energy, the 
renormalization group (RG) running of the effective neutrino
mass matrix has to be taken into account.   

\subsection{Renormalization Group Corrections}\label{sec:RGRunning}
For type I models with sequential dominance, the running of the mixing angles 
is generically small \cite{King:2000hk} 
since the mass scheme is strongly hierarchical.  
When the neutrino mass scale is lifted, e.g.\ via a type II upgrade, 
a careful treatment of
the RG running of the neutrino parameters, 
including the energy ranges between and above the see-saw
scale \cite{King:2000hk,betakappa}, is required. 
For convenient estimates of the running below the see-saw scales, 
the approximate analytical formulae for 
the running of the parameters \cite{Antusch:2003kp} 
can be used. Dependent on $\tan \beta$ in the MSSM, on the size of the 
neutrino Yukawa couplings and on the neutrino mass scale, the RG effects can  
be sizable or cause only small corrections.\footnote{A complete list of  
references for the $\beta$-functions of the neutrino mass operator 
can be found in \cite{Antusch:2003kp}.}

\subsection{Dirac and Majorana CP Phases and Neutrinoless Double Beta Decay} 
At present, the CP phases in the lepton sector are unconstraint by experiment. 
In type I see-saw models based on sequential dominance, there is no 
restriction on them from a theoretical point of view.  
The type-II-upgrade scenario however predicts that all observable 
CP phases, i.e.~the Dirac CP phase $\delta$ relevant for neutrino oscillations 
and the Majorana CP phases $\beta_2$ and $\beta_3$,  
become small as the neutrino mass scale increases. 

The key process for measuring the neutrino mass scale could be
neutrinoless double beta decay. The decay rates depend on an effective
Majorana mass defined by $\< m_\nu \>= \left| \sum_i
(U_\mathrm{MNS})_{1i}^2 \, m_i \right|$.  Future experiments which are
under consideration at present might increase the sensitivity to $\<
m_\nu \>$ by more than an order of magnitude. 
For type I models with
sequential dominance, which have a hierarchical mass scheme, $\< m_\nu
\>$ can be very small, below the accessible sensitivity.

For models where the neutrino mass scale is lifted via a type II upgrade 
\cite{Antusch:2004xd}, there
is a close relation between the neutrino mass scale, i.e.\ the mass of
the lightest neutrino and $\< m_\nu \>$. Since the CP phases are
small, there can be no significant cancellations in $\< m_\nu \>$.  This
implies that the effective mass for neutrinoless double beta decay is
approximately equal to the neutrino mass scale
$\< m_\nu \>\approx m^{\mathrm{II}}$
and therefore
neutrinoless double beta decay will be observable in the next round of
experiments if the neutrino mass spectrum is partially degenerate.

\subsection{Theoretical expectations for the Mixing Angles}
In order to discriminate between models, precision measurements of the 
neutrino mixing angles have the potential to play an important role.

One important parameter is the value of the mixing angle
$\theta_{13}$, which is at present only bounded from above to be
smaller than approximately $13^\circ$.  In the type I sequential
dominance case, the mixing angle $\theta_{13}$ is typically of the
order ${\cal O}(m^\mathrm{I}_2/m^{\mathrm{I}}_3)$.  In the
type-II-upgrade scenario this ratio decreases with increasing neutrino
mass scale and is smaller than $\approx 5^\circ$ for partially
degenerate neutrinos even if it was quite large in the
type I limit. Sizable RG corrections, which are usually expected for
partially degenerate neutrinos, are suppressed in the type-II-upgrade
scenario due to small CP phases $\beta_2,\beta_3$ and $\delta$
\cite{Antusch:2003kp}.

Another important parameter is $\theta_{23}$. Its present best-fit value is
close to $45^\circ$, however comparably large deviations are experimentally
allowed as well.   
With sequential dominance, we expect minimal deviations of  
$\theta_{23}$ from $45^\circ$ of the order 
${\cal O}(m^\mathrm{I}_2/m^{\mathrm{I}}_3)$, which could be observed 
by future long-baseline experiments in the type I 
see-saw case.\footnote{For sensitivities of future long-baseline experiments 
for measuring deviation from $\theta_{23} =45^\circ$ and their potential for
discriminating between models, see \cite{Antusch:2004yx}.} 
In the type II upgraded version, the corrections can be significantly smaller
since the ratio $m^\mathrm{I}_2/m^{\mathrm{I}}_3$ decreases with increasing
neutrino mass scale \cite{Antusch:2004xd}. For large $\tan \beta$ in the MSSM, the major source for 
the corrections can be RG effects \cite{Antusch:2003kp}, which are
un-suppressed for small CP phases.

\subsection{Minimal Lepton Flavour Violation} 

At leading order in a mass insertion approximation
the branching fractions of LFV processes are given by
\footnote{The mass insertion approximation given in
  Eq.\ref{eq:BR(li_to_lj)} is for illustrative purposes only.
The conclusions quoted below from \cite{Blazek:2002wq}
do not rely on this approximation.}
\beq
{\rm BR}(l_i \rightarrow l_j \gamma)\approx
        \frac{\alpha^3}{G_F^2}
        f(M_2,\mu,m_{\tilde{\nu}})
        |m_{\tilde{L}_{ij}}^2|^2  \tan ^2 \beta
    \label{eq:BR(li_to_lj)}\; ,
\eeq
where $l_1=e, l_2=\mu , l_3=\tau$,
and where the off-diagonal slepton doublet mass squared is given
in the leading log approximation (LLA) by
\beq
m_{\tilde{L}_{ij}}^{2(LLA)}
\approx -\frac{(3m_0^2+A_0^2)}{8\pi ^2}C_{ij}\; .
\label{lla}
\eeq
With sequential dominance, using the notation of Eqs.\ref{seq1},\ref{dirac}, 
the leading log coefficients relevant for
$\mu \rightarrow e\gamma$ and $\tau \rightarrow \mu \gamma$
are given approximately as
\bea
C_{21} & = & ab\ln \frac{M_U}{X} +de\ln \frac{M_U}{Y}\; , \nonumber \\
C_{32} & = & bc\ln \frac{M_U}{X} +ef\ln \frac{M_U}{Y}\; .
\label{C2131}
\eea
From Table~\ref{table1} and Eq.\ref{C2131} it can be seen
which types of SD will lead to large rates for 
$\mu \rightarrow e\gamma$ and $\tau \rightarrow \mu \gamma$.
For example the results for HSD show a large rate for
$\tau \rightarrow \mu \gamma$ which is the characteristic expectation
of lop-sided models in general \cite{Blazek:2001zm} and HSD in
particular. A global analysis of LFV has been performed in the constrained
minimal supersymmetric standard model (CMSSM) for the case
of sequential dominance, focussing on the two cases
of HSDa and LSDa \cite{Blazek:2002wq}. The results in \cite{Blazek:2002wq}
are based on an exact calculation, and
the error incurred compared to the LLA study \cite{Lavignac:2002gf}
can be as much as 100\%.
For LSDa $\tau \rightarrow \mu \gamma$
is well below observable values. Therefore $\tau \rightarrow \mu
\gamma$ provides a good discriminator between the HSDa and LSDa types
of dominance. In \cite{Blazek:2002wq} it is shown that 
the rate for $\mu \rightarrow e \gamma$ may determine the order
of the sub-dominant neutrino Yukawa couplings in the flavour basis.

\subsection{Leptogenesis}

Leptogenesis and lepton flavour violation are important indicators
which can help to resolve the ambiguity of right-handed neutrino
masses in Table 1. In the LSD and HSD cases of sequential dominance
leptogenesis has been studied
with some interesting results \cite{Hirsch:2001dg}. 
In general successful
leptogenesis for such models requires the mass of the lightest
right-handed neutrino to be quite high, and generally to exceed the
gravitino constraints if supersymmetry is assumed.
However, putting this to one side for the moment,
interesting links between the phase relevant for leptogenesis
and the phase $\delta$ measurable in neutrino oscillation experiments
have been made. The precise link depends on how many ``texture'' 
zeroes are assumed to be present in the neutrino Dirac mass matrix.
For example if two texture zeroes are assumed then there is a
direct link between $\delta$ and the leptogenesis phase,
with the sign of $\delta$ being predicted from the fact that we are
made of matter rather than antimatter. On the other hand if
only the physically motivated texture zero in the 11 entry of the
Dirac mass matrix is assumed, then the link is more indirect
\cite{King:2002qh}.

More generally in three right-handed neutrino models with sequential
dominance, if the dominant right-handed neutrino is the lightest
one (LSD) then the washout parameter $\tilde{m_1}\sim O(m_3)$, 
which is rather too large compared to the optimal value
of around $10^{-3}$ eV,
while if the dominant right-handed neutrino is either the
intermediate one or the heaviest one then one finds
$\tilde{m_1}\sim O(m_2)$ or arbitrary $\tilde{m_1}$, which can be 
closer to the desired value \cite{Hirsch:2001dg}.

\section{Discussion and Conclusions}
Neutrino masses and mixings are now established experimental
phenomena which must be included in some extended version
of the Standard Model. The simplest mechanism for describing
small neutrino masses is the see-saw mechanism, however the
simultaneous appearance of hierarchical neutrino masses
and two large mixing angles is not natural in the see-saw
mechanism. The simplest solution to this difficulty is to assume 
sequential dominance which has been the subject of this review.

We have reviewed the mechanism of sequential right-handed neutrino 
dominance which was proposed in the
framework of the type I see-saw mechanism to account for bi-large
neutrino mixing and a neutrino mass hierarchy in a natural way. 
We have discussed how sequential dominance may also be applied to the
right-handed charged leptons, which alternatively 
allows bi-large lepton mixing in the charged lepton sector. We reviewed how
such sequential dominance models may be upgraded to include type II
see-saw contributions, resulting in a partially degenerate neutrino
mass spectrum with bi-large lepton mixing arising from sequential
dominance. We also saw that the use of sequential 
dominance is ideally suited to GUTs and family symmetry models,
and mentioned some examples of such models.
We also pointed out the interesting case
where sequential dominance reduces effectively to the
case of two right-handed neutrinos, and mentioned some interesting
cosmological applications that have recently been proposed
such as sneutrino inflation.

We also reviewed some 
phenomenological consequences of type I see-saw models with 
sequential dominance and their type II upgrades for the low energy 
neutrino parameters and high-energy mechanisms as leptogenesis and 
minimal lepton flavour violation, both of which can be probes of
different types of sequential dominance. While RG effects are
expected to be quite small for type I sequential dominance,
they become increasingly important for the type II upgrade
sequential dominance as the neutrino mass scale increases. 
We noted that neutrinoless double beta decay is practically
unobservable in type I sequential dominance, but may well be observed
in the next round of experiments in the type II upgrade sequential
models if the neutrino masses are partially degenerate.
We saw that both $\theta_{13}$ and the correction to 
$\theta_{23}$ are controlled by the ratio 
$m^\mathrm{I}_2/m^{\mathrm{I}}_3$ which decreases with increasing
neutrino mass scale, with interesting consequences.

Given the simplicity and naturalness of sequential dominance,
we expect it to continue to be used and exploited ubiquitously in the future.

\section*{Acknowledgements}
We acknowledge support from the PPARC grant PPA/G/O/2002/00468.

\appendix

\section{Our Conventions}\label{conventions}
For the mass matrix of the charged leptons $m_{\mathrm{E}}=Y_e v_d$, where 
$v_d = \< H_\mathrm{d}\>$,  
defined by 
$\mathcal{L}_e=-m_{\mathrm{E}} \overline e^f_{\mathrm{L}} e^f_{\mathrm{R}}$ + h.c. 
and for the neutrino mass
matrix $m^\nu_{LL}$, 
the change from flavour basis to mass
eigenbasis can be performed with the unitary diagonalization matrices 
$U_{e_\mathrm{L}},U_{e_\mathrm{R}}$ and 
$U_{\nu_\mathrm{L}}$ by  
\begin{eqnarray}\label{eq:DiagMe}
U_{e_\mathrm{L}} \, m_{\mathrm{E}} \,U^\dagger_{e_\mathrm{R}} =
\left(\begin{array}{ccc}
\!m_e&0&0\!\\
\!0&m_\mu&0\!\\
\!0&0&m_\tau\!
\end{array}
\right)\! , \quad
U_{\nu_\mathrm{L}} \,m^\nu_{\mathrm{LL}}\,U^T_{\nu_\mathrm{L}} =
\left(\begin{array}{ccc}
\!m_1&0&0\!\\
\!0&m_2&0\!\\
\!0&0&m_3\!
\end{array}
\right)\! .
\end{eqnarray}  
The MNS matrix is then given by
\begin{eqnarray}
U_{\mathrm{MNS}} = U_{e_\mathrm{L}} U^\dagger_{\nu_\mathrm{L}}\; .
\end{eqnarray}
We use the parameterization 
$
U_{\mathrm{MNS}} = R_{23} U_{13} R_{12} P_0
$ 
with $R_{23}, U_{13}, R_{12}$ and $P_0$ being defined as 
\begin{align}\label{eq:R23U13R12P0}
R_{12}:=
\left(\begin{array}{ccc}
  c_{12} & s_{12} & 0\\
  -s_{12}&c_{12} & 0\\
  0&0&1\end{array}\right)
  , \:&
\quad U_{13}:=\left(\begin{array}{ccc}
   c_{13} & 0 & s_{13}e^{-i\delta}\\
  0&1& 0\\
  - s_{13}e^{i\delta}&0&c_{13}\end{array}\right)  ,  \nonumber \\ 
R_{23}:=\left(\begin{array}{ccc}
 1 & 0 & 0\\
0&c_{23} & s_{23}\\
0&-s_{23}&c_{23}
 \end{array}\right)
  , \:&
 \quad P_0:= 
 \begin{pmatrix}
 1&0&0\\0&e^{i\beta_2}&0\\0&0&e^{i\beta_3}
  \end{pmatrix}  
\end{align}
and where $s_{ij}$ and $c_{ij}$ stand for $\sin (\theta_{ij})$ and $\cos
(\theta_{ij})$, respectively. 
The matrix $P_0$ contains the possible Majorana 
phases $\beta_2$ and $\beta_3$.  
 $\delta$ is the Dirac CP phase relevant for neutrino oscillations.

\end{document}